\documentclass{jpp}
\usepackage{graphicx}

\usepackage[utf8]{inputenc}
\usepackage[T1]{fontenc}
\usepackage{amsmath}

\usepackage{color}

\newcommand{\chib}{\bar{\chi}}
\newcommand{\bn}{\begin{equation}}
\newcommand{\bea}{\begin{eqnarray*}}
\newcommand{\eea}{\end{eqnarray*}}
\newcommand{\en}{\end{equation}}
\newcommand{\lang}{\left\langle}
\newcommand{\rang}{\right\rangle}
\newcommand{\ap}{b}
\newcommand{\vTa}{v_{Ta}}
\newcommand{\vpar}{v_\parallel}
\newcommand{\vperp}{v_\perp}
\newcommand{\Tau}{{\mathcal T}}
\newcommand{\Eps}{{\mathcal E}}
\newcommand{\wt}{t}

\shorttitle{Part II}
\shortauthor{G. G. Plunk and P. Helander}

\title{Energetic bounds on gyrokinetic instabilities. Part II. Modes of optimal growth.}
\author{G. G. Plunk\aff{1}
  \corresp{\email{gplunk@ipp.mpg.de}},
  \and P. Helander\aff{1}}

\affiliation{\aff{1}Max-Planck-Institut für Plasmaphysik, 17491 Greifswald, Germany}

\begin{document}

\maketitle

\begin{abstract}
We introduce modes of instantaneous optimal growth of free energy for the fully electromagnetic gyrokinetic equations.  We demonstrate how these ``optimal modes'' arise naturally from the free energy balance equation, allowing its convenient decomposition, and yielding a simple picture of energy flows.  Optimal modes have a number of other favorable features, such as their low-dimensionality, efficiency of computation, and the fact that their growth rates provide a rigorous and ``tight'' upper bound on both the nonlinear growth rate of energy, and the linear growth rate of traditional gyrokinetic (normal mode) instabilities.  We provide simple closed form solutions for the optimal growth rates in a number of asymptotic limits, and compare these with our previous bounds.
\end{abstract}

\section{Introduction}

This is the second paper in a series, in which we develop a linear and nonlinear stability theory based on gyrokinetic energy balance.  Whereas the first paper emphasizes simple and rigorous upper bounds, this second paper shifts focus to tightening these bounds (making them exactly realizable under the right initial conditions), fully generalizing their validity (allowing fully electromagnetic fluctuations including non-zero $\delta B_\parallel$), and introducing the notion of a complete orthogonal set of ``optimal modes'' associated with these bounds.

Gyrokinetic stability analysis serves as the foundation of turbulence and transport theory for magnetic fusion devices, with linear calculations being the starting point for predicting turbulence intensity and other properties.  Such calculations are also the main ingredient of mixing-length estimates \citep{horton-rev}, quasi-linear theory \citep{Bourdelle_2015}, and other (indeed, perhaps all) transport models.  These linear instabilities are typically understood (with some notable exceptions, {\em e.g.} \cite{Hatch_2016}) as ``normal modes'', i.e. eigenmodes of the linearized gyrokinetic equation, whose time dependence takes the form $\sim \exp(-i\omega t)$, where $\Imag[\omega] = \gamma_\mathrm{L}$ is the growth rate.  When numerically computing gyrokinetic normal modes, it is most common to solve an initial value problem, and terminate the computation only when the solution is found to fit the exponential form.  This technique yields the mode of largest growth rate for a given wave number.

This paper introduces a different kind of gyrokinetic mode, one that realizes optimal growth of energy at an instant in time (actually, the idea was proposed first in the gyrokinetic context by \citet{landreman_plunk_dorland_2015} but not calculated explicitly there).  These modes have been studied extensively in fluid turbulence, where they are called ``instantaneous optimal perturbations'', and are found as a limit of a more general class of modes that achieve optimal growth of energy over a finite time interval \citep{farrell-generalized-stability-theory}.%  We will call them ``optimal modes'' for short.

To illustrate the essential idea of such modes, free from the complications of full gyrokinetics, let us consider a simple schematic equation for a linearized dynamical system with a complex state variable $\psi(t,\ell)$, where $t$ is time and $\ell$ is a continuous space variable,

\begin{equation}
    \frac{\partial \psi}{\partial t} = \cal{L} \psi,\label{proto-system}
\end{equation}
with $\cal{L}$ representing some linear operator, whose eigenmodes are precisely the ``normal modes'' described above.  Defining an inner product (involving, {\em e.g.}, a suitable average over $\ell$), we obtain an ``energy'' evolution equation for $\left( \psi,  \psi \right) = ||\psi||^2$

\begin{equation}
    \frac{d ||\psi||^2}{d t} = \left( \psi,{\cal H} \psi\right),\label{proto-optimal-eqn}
\end{equation}
where ${\cal H} = \cal{L} + \cal{L}^\dagger$ is a Hermitian linear operator, with the adjoint operator $\cal{L}^\dagger$ defined as that for which $\left( \psi_1,{\cal L} \psi_2\right) = \left( {\cal L}^{\dagger} \psi_1, \psi_2\right)$ holds for arbitrary $\psi_1$ and $\psi_2$ in the Hilbert space defined by the inner product.\footnote{We mean ``Hilbert space'' in the sense of a physicist, {\em i.e.} $\psi$ may have dependence proportional to the Dirac delta function when needed \citep{shankar-book}.}  From this we can immediately surmise that the eigenmodes of ${\cal H}$, {\em i.e.} $\psi_n$ such that ${\cal H}\psi_n = \lambda_n \psi_n$, can be used to characterize the energy growth of the system, as their orthogonality reduces the right hand side of this equation to the sum of the squares of eigenmode amplitudes, times their eigenvalues.  These are precisely the ``optimal modes'' described above.  Writing $\psi = \sum_n c_n \psi_n$, and taking $||\psi_n||^2 = 1$, we have%\footnote{The spectrum of modes is generally not discrete in the case of gyrokinetics, and so the analagous expression will need to be replaced with some combination of integration and summation.}
\begin{equation}
    \sum_n \frac{d |c_n|^2}{dt} = \sum_n \lambda_n  |c_n|^2.\label{proto-energy-balance}
\end{equation}
Obviously, the eigenmode with the largest positive value achieves optimal growth, {\it i.e.} maximizes $||\psi||^{-2} d ||\psi||^2/dt $.  

Note that ``instantaneous optimal perturbations'' achieve optimal growth only for an instant.  That is, if the system is initialized to be an optimal solution, the observed growth of energy will generally only match the theoretical value initially, and the solution, left undisturbed by nonlinear physics, will evolve over a certain timescale toward a normal mode described above.  Clearly, if this timescale is much longer than the nonlinear decorrelation rate of the turbulence, then the growth of normal modes cannot be expected to be a useful measure, and an alternative measure, such as the optimal rate, may be better.  Furthermore, there are cases where turbulence arises when no unstable normal modes exist at all, this being the original motivation for introducing optimal modes.  For so-called ``sub-critical'' turbulence, one needs a measure to characterize the transient linear growth of fluctuations.  It is not difficult to see that modes of positive instantaneous optimal growth, one such measure, are {\em necessary} for sub-critical turbulence to exist \citep{del-sole_necessity}.  Indeed, from Eqn.~\ref{proto-energy-balance}, the energy of the system can only ever increase if eigenmodes exist with $\lambda_n > 0$.

Even in cases where growth rate of normal modes is found to be a suitable measure of instability, we note that the analysis of optimal modes provides a rigorous upper bound to that growth: if $\gamma$ is an eigenvalue of ${\cal L}$ in the system above, then $\gamma \leq \max_n \lambda_n/2$.

For gyrokinetics, we can point to several further advantages of optimal modes.  The first, as we will see, is the drastic reduction in the dimensionality of the problem, as compared to normal mode analysis, for which the kinetic distribution of each particle species requires two velocity variables to be accounted for, in addition to time and space.  The complete analysis of optimal modes, in contrast, requires only a few velocity moments to be computed, eliminating the vast majority of the complexity of velocity space.  Explicit time evolution, required in initial value computations, is also avoided, and the modes we find are ``local'' in position space, {\em i.e.} the field-line following coordinate becomes merely a parameter of the theory.  Furthermore, the singular nature of modes in kinetic theory ({\em e.g.} Case-Van Kampen modes, {\em etc.}) is evaded, and the optimal modes form a complete and orthogonal basis for the entire space of gyrokinetic fluctuations.  This gives a natural way to decompose energy balance (see Eqn.~\ref{proto-energy-balance}), and simplifies the analysis of energy flows in the turbulent steady state -- the turbulence is excited by unstable modes (with $\lambda_n > 0$) and damped by stable ones (with $\lambda_n < 0$).  Optimal mode theory,  following directly from energy balance, gives a universal bound on the full ``zoo'' of gyrokinetic instabilities, in contrast to traditional linear (normal) mode analysis, which requires detailed, separate analysis of different cases.  Optimal modes therefore give a unified picture of the space of instabilities.

As we will see, many useful cases of optimal growth can be evaluated with simple closed-form expressions, without the need for numerical computation; when such computation is needed, its cost is trivial.  In short, we will show that optimal modes can be viewed as a theoretically transparent and computationally efficient alternative (or complement) to normal modes in gyrokinetics.

\section{Definitions and gyrokinetic free energy balance}

We are interested in finding the solutions that optimize the growth of a certain measure of fluctuation energy.  The choice that we have made so far in this series of articles is the gyrokinetic Helmholtz free energy, commonly referred to as simply ``free energy''.  This measure has the advantage of being a ``nonlinear invariant'', {\it i.e.} conserved under nonlinear interactions, and also having a satisfying thermodynamic interpretation -- it  only diminishes under the action of collisions.  These facts are well known and spelled out in part I of this series.  We can therefore begin directly with the energy balance equation, which, neglecting collisions, reads

\begin{equation}
    \frac{d}{d t}\sum_{\bf k} H = 2 \sum_{\bf k} D,\label{energy-balance}
\end{equation}
where the perpendicular wavenumber is ${\bf k} = {\bf k}_\perp = k_\psi \bnabla \psi + k_\alpha \bnabla \alpha$, in terms of magnetic coordinates $\psi$ and $\alpha$ such that the equilibrium magnetic field is ${\bf B} = \bnabla \psi \times \bnabla \alpha$.  The drive term $D$ is

\begin{equation}
    D({\bf k}, t) = {\rm Im} \; \sum_a  e_a \lang \int g_{a, \bf k} \omega_{\ast a}^T \chib^\ast_{a,\bf k} d^3v \rang, 
	\label{D-eqn}
\end{equation}
and the free energy is

\begin{equation}
    H({\bf k},t)  = \sum_a \lang  T_a \int \frac{|g_{a, \bf k}|^2}{F_{a0}} d^3v - \frac{n_a e_a^2}{T_a} |\delta \phi_{\bf k}|^2 \rang
	+ \lang \frac{| \delta {\bf B}_{\bf k} |^2}{\mu_0} \rang,\label{H-eqn}
\end{equation}
where $| \delta {\bf B}_{\bf k} |^2 = | k_\perp \delta A_{\|\bf k} |^2 + | \delta B_{\|\bf k} |^2$, and we define the other notation as follows.  The space average is defined as\footnote{Our results also hold for a more general definition of the space average, as discussed in \citet{helander-plunk-jpp-part-1}.}

\begin{equation}
    \lang \cdots \rang = \lim_{L\rightarrow \infty} \int_{-L}^L (\cdots ) \frac{dl}{B} \bigg\slash
	\int_{-L}^L \frac{dl}{B}.
\end{equation}
The diamagnetic frequencies are

	$$ \omega_{\ast a} = \frac{k_\alpha T_a}{e_a} \frac{d \ln n_a}{d \psi}, $$
	$$ \omega_{\ast a}^T = \omega_{\ast a} 
	\left[1 + \eta_a \left( \frac{m_a v^2}{2 T_a} - \frac{3}{2} \right)\right]. $$
The gyro-averaged electromagnetic potential is

	$$ \chib_{a \bf k} 
	= J_0 \left( \frac{k_\perp v_\perp}{\Omega_a} \right)\left( \delta \phi_{{\bf k}} - v_\| \delta A_{\| {\bf k}} \right)
	+ J_1 \left( \frac{k_\perp v_\perp}{\Omega_a} \right) \frac{v_\perp}{k_\perp} \delta B_{\| \bf k}, $$
and $J_0$ and $J_1$ are Bessel functions. Field perturbations are given by 
\begin{eqnarray}
    \sum_a \frac{n_a e_a^2}{T_a} \;\delta \phi_{\bf k} = \sum_a e_a \int g_{a, {\bf k}} J_{0a} d^3v, 
	\label{field1}\\
	\delta A_{\| {\bf k}} = \frac{\mu_0}{k_\perp^2} \sum_a e_a \int v_\| g_{a, {\bf k}} J_{0a} d^3v, 
	\label{field2}\\
	\delta B_{\| {\bf k}} = - \frac{\mu_0}{k_\perp} \sum_a e_a \int v_\perp g_{a, {\bf k}} J_{1a} d^3v,
	\label{field3}
\end{eqnarray}%$ \lambda_a = {n_a e_a^2}/{T_a}$ and 
where we define $J_{na} = J_n(k_\perp v_\perp / \Omega_a)$.  Henceforth consider a single fixed value of $\bf k$ and therefore suppress the $\bf k$-subscripts.

Note that we define the free energy as twice that which appears in some other publications, but this has no significant effect on the analysis, as the 2 drops out of the optimization problem, upon division by $H$.

\section{Modes of optimal instantaneous growth}

To recast Eqn.~\ref{energy-balance} in the form of Eqn.~\ref{proto-optimal-eqn} requires first identifying the state variable(s) and inner product.  We take the state to be the set of distributions functions, given by the vector ${\bf g}$ with components $g_{a}$, and define the inner product of two states ${\bf g}_1$ and ${\bf g}_2$ as

\begin{equation}
    ({\bf g}_1, {\bf g}_2) =  \sum_{a}\lang T_a \int  \frac{g_{a1}^*g_{a2}}{F_{a0}} d^3v \rang.
\end{equation}
%The inner product is needed to establish the sense in which the linear operators found are Hermitian.

Comparing with Eqn.~\ref{H-eqn}, we note that the free energy $H$ is not the Euclidean norm in these variables (though it is positive-definite in ${\bf g}$ for non-zero wavenumber; see \citet{helander-plunk-jpp-part-1}).  Although it is possible to transform to new state variables, {\em i.e.} $\tilde{\bf g}$ such that $H = ||\tilde{{\bf g}}||^2$, we find it is more convenient to take another approach at this stage, namely to formulate our problem in variational terms.  That is, we extremize the ratio 

\begin{equation}
    \Lambda = D/H\label{Lambda-eqn}
\end{equation}
over the space of distribution functions ${\bf g}$.  Note also that, as is easily verified, normal modes satisfy $\gamma_L = D/H$, so that we have the bound on linear gyrokinetic instabilities:

\begin{equation}
    \gamma_L \leq \max_{g_{a}} \Lambda.\label{eq:linear-bound}
\end{equation}
Variation of Eqn.~\ref{Lambda-eqn} leads to the condition

\begin{equation}
    \frac{\delta D}{\delta g_{a}} - \Lambda \frac{\delta H}{\delta g_{a}} = 0.\label{variational-problem}
\end{equation}
Note that $\Lambda$, which according to Eqn.~\ref{energy-balance} corresponds to {\em half} the growth rate of free energy, can be interpreted as the Lagrange multiplier whose role is to hold $H$ fixed.  

Using Eqns.~\ref{D-eqn} and \ref{H-eqn}, we can evaluate Eqn.~\ref{variational-problem} to obtain (see Appendix \ref{variation-appx})

\begin{equation}
   \Lambda \sum_{\ap} {\cal H}_{a \ap} g_{\ap} = \sum_{\ap} {\cal D}_{a \ap} g_{\ap},\label{Kinetic-eigenprob}
\end{equation}
where ${\cal H}$ and ${\cal D}$ are, respectively, a purely real and purely imaginary Hermitian linear operators on the space of distribution functions, given as follows:

\begin{equation}\label{cal-H-eqn}
    {\cal H}_{a \ap} g_{\ap} = \delta_{a,\ap} g_{\ap} + \frac{F_{a0}}{n_a T_a} \frac{1}{n_b} \int d^3v^\prime g_{\ap}^\prime \left[ -\wt_{a} \wt_{\ap} \psi_{1a}\psi_{1b}^\prime + \varepsilon_{a} \varepsilon_{\ap}(\psi_{3a}\psi_{3b}^\prime + \psi_{5a}\psi_{5b}^\prime) \right],
\end{equation}
noting that the species label $\ap$ is not to be confused with the argument of the Bessel functions, and $\delta_{a,\ap}$ is the discrete delta function.  The second operator is given by

\begin{align}\label{cal-D-eqn}
    {\cal D}_{a \ap} g_{\ap} = &\frac{i}{2}\frac{F_{a0}}{n_a T_a}  \frac{1}{n_{\ap}} \int d^3v^\prime g_{\ap}^\prime \left[ \right.\nonumber\\ 
    &\omega_{\ast a}\left(1-3\eta_a/2\right)\left( \wt_{a} \wt_{\ap} \psi_{1a}\psi_{1\ap}^\prime - \varepsilon_{a} \varepsilon_{\ap}\psi_{3a}\psi_{3\ap}^\prime - \varepsilon_{a} \varepsilon_{\ap}\psi_{5a}\psi_{5\ap}^\prime \right) \nonumber\\
    - &\omega_{\ast \ap}\left(1-3\eta_{\ap}/2\right) \left( \wt_{a} \wt_{\ap} \psi_{1a}\psi_{1\ap}^\prime - \varepsilon_{a} \varepsilon_{\ap}\psi_{3a}\psi_{3\ap}^\prime - \varepsilon_{a} \varepsilon_{\ap}\psi_{5a}\psi_{5\ap}^\prime \right)\nonumber\\
    + &\omega_{\ast a} \eta_a \left( \wt_{a} \wt_{\ap} \psi_{2a}\psi_{1\ap}^\prime - \varepsilon_{a} \varepsilon_{\ap}\psi_{4a}\psi_{3\ap}^\prime - \varepsilon_{a} \varepsilon_{\ap}\psi_{6a}\psi_{5\ap}^\prime \right)\nonumber\\
    - &\omega_{\ast \ap} \eta_{\ap} \left( \wt_{a} \wt_{\ap} \psi_{1a}\psi_{2\ap}^\prime - \varepsilon_{a} \varepsilon_{\ap}\psi_{3a}\psi_{4\ap}^\prime - \varepsilon_{a} \varepsilon_{\ap}\psi_{5a}\psi_{6\ap}^\prime \right)\left. \right],
\end{align}
where primes denote evaluation at $v^\prime$, and we define
\begin{eqnarray}
    \wt_{a} &= &e_a n_a \left( \sum_{a^\prime} \frac{n_{a^\prime} e_{a^\prime}^2}{T_{a^\prime}}\right)^{-1/2},\\
    \varepsilon_{a} &=  &e_a n_a \left(\frac{\sqrt{\mu_0} \vTa }{k_\perp}\right) = \mathrm{sgn}(e_a)\sqrt{n_a T_a \beta_a/b_a},
\end{eqnarray}%$\mathrm{sgn}(q_a) = \pm 1$ depending on the sign of $q_a$, and
where $\mathrm{sgn}(e_a) = \pm 1$ gives the sign of $e_a$, $b_a = k_\perp^2 m_a T_a/ (e_a^2 B^2)$, the plasma beta of species $a$ is $\beta_a = 2\mu_0 n_a T_a/B^2$, and its thermal velocity is denoted $\vTa = \sqrt{2 T_a/m_a}$.  We also introduce velocity dependent functions that are needed for forming the relevant moments
% \begin{subeqnarray}
%     \psi_{1a} &= &J_{0a}\\
%     \psi_{2a} &= &\frac{v^2}{\vTa^2}J_{0a}\\
%     \psi_{3a} &= &\frac{\vpar}{\vTa} J_{0a}\\
%     \psi_{4a} &= &\frac{\vpar v^2}{\vTa^3}J_{0a}\\
%     \psi_{5a} &= &\frac{\vperp}{\vTa}J_{1a}\\
%     \psi_{6a} &= &\frac{\vperp v^2}{\vTa^3}J_{1a}
% \end{subeqnarray}

\begin{equation}
\begin{tabular}{l l l}
 $\psi_{1a} = J_{0a},$ & $\psi_{2a} = \frac{v^2}{\vTa^2}J_{0a},$ & $\psi_{3a} = \frac{\vpar}{\vTa} J_{0a},$ \\
 $\psi_{4a} = \frac{\vpar v^2}{\vTa^3}J_{0a},$ & $\psi_{5a} = \frac{\vperp}{\vTa}J_{1a},$ & $\psi_{6a} = \frac{\vperp v^2}{\vTa^3}J_{1a}.$
\end{tabular}
\end{equation}

\subsection{Moment form of Eqn.~\ref{Kinetic-eigenprob}}

Eqn.~\ref{Kinetic-eigenprob} describes an eigenproblem whose solutions form a complete orthogonal basis for the space of distribution functions $g_{a}$.  This is a large space, but as we will see, the non-trivial solution space of \ref{Kinetic-eigenprob} is actually quite small.  There are a number of reasons for this, but the first and most important is that only a small set of velocity moments appear in this equation.  We can identify six dimensionless moments for each species, defined as

\begin{equation}
    \kappa_{n a} = \frac{1}{n_a} \int d^3v \;\psi_{n a} g_a.
\end{equation}
We further note that, due to the summation over species involved in the computation of the electromagnetic fields and free energy balance, the velocity moments $\kappa_n$ appear in particular linear combinations.  Thus there is an additional dimensional reduction corresponding to the number of species $N_s$, {\em i.e.} $6 N_s \rightarrow 6$.  This is achieved with the help of the following barred variables

\begin{equation}\label{kappa-bar-eqns}
\begin{tabular}{l  l}
 $\bar{\kappa}_1 = \sum_{a} \wt_{a} \kappa_{1 a}$, & $\quad\bar{\kappa}_2 = \sum_{a} \wt_{a} \bar{\omega}_{\ast a}\left(\left(1-3\eta_{a}/2\right)\kappa_{1 a} + \eta_{a} \kappa_{2 a} \right)$, \\
 $\bar{\kappa}_3 = \sum_{a} \varepsilon_{a} \kappa_{3 a}$, & $\quad\bar{\kappa}_4 = \sum_{a} \varepsilon_{a} \bar{\omega}_{\ast a}\left(\left(1-3\eta_{a}/2\right)\kappa_{3 a} + \eta_{a} \kappa_{4 a} \right)$, \\
 $\bar{\kappa}_5 = \sum_{a} \varepsilon_{a} \kappa_{5 a}$, & $\quad\bar{\kappa}_6 = \sum_{a} \varepsilon_{a} \bar{\omega}_{\ast a}\left(\left(1-3\eta_{a}/2\right)\kappa_{5 a} + \eta_{a} \kappa_{6 a} \right)$, \\
\end{tabular}
\end{equation}
where we introduce the normalized frequency $\bar{\omega}_{\ast a} = \omega_{\ast a}/\omega_{\ast}$, with $\omega_{\ast}$ an arbitrary reference value. Evaluating the sum over species in Equation \ref{Kinetic-eigenprob} yields

\begin{multline}
  \frac{\Lambda}{\omega_{*}} \left( g_{a} + \frac{F_{a0}}{n_a T_a} \left[ -\wt_{a} \psi_{1a}\bar{\kappa}_1 + \varepsilon_{a} (\psi_{3a}\bar{\kappa}_3 + \psi_{5a}\bar{\kappa}_5) \right]\right)  \\
  = \frac{i}{2}\frac{F_{a0}}{n_a T_a}  \left[ \right. 
    \bar{\omega}_{\ast a}\left(1-3\eta_a/2\right)\left( \wt_{a} \psi_{1a}\bar{\kappa}_1 - \varepsilon_{a} \psi_{3a}\bar{\kappa}_3 - \varepsilon_{a} \psi_{5a}\bar{\kappa}_5 \right) \\
    +\bar{\omega}_{\ast a} \eta_a \left( \wt_{a} \psi_{2a}\bar{\kappa}_1 - \varepsilon_{a} \psi_{4a}\bar{\kappa}_3 - \varepsilon_{a} \psi_{6a}\bar{\kappa}_5 \right)\\
    -\wt_{a}\psi_{1a}\bar{\kappa}_2 + \varepsilon_{a} \psi_{3a} \bar{\kappa}_4 + \varepsilon_{a} \psi_{5a} \bar{\kappa}_6 \left. \right].\label{Reduced-kinetic-eigenprob}
\end{multline}

To simplify the problem further, we can take moments of Equation \ref{Reduced-kinetic-eigenprob} and obtain a closed linear system for $\bar{\kappa}_n$.  We first write Equation \ref{kappa-bar-eqns} as

\begin{equation}
    \bar{\kappa}_m = \sum_{n,\ap} c_{mn}^{(\ap)} \kappa_{n \ap}.\label{kappa-bar-c-eqn}
\end{equation}
Now taking moments of \ref{Reduced-kinetic-eigenprob}, and summing over species, using Equation \ref{kappa-bar-c-eqn}, we obtain
\begin{multline}%\frac{c_{mn}^{(a)}}{n_a T_a}
    \frac{\Lambda}{\omega_{\ast}} \left( \bar{\kappa}_m + \sum_{a, n} \left\{-\Tau_{mn}^{(a)} X_{1n}^{(a)} \bar{\kappa}_1 + \Eps_{mn}^{(a)} (X_{3n}^{(a)} \bar{\kappa}_3+X_{5n}^{(a)} \bar{\kappa}_5) \right\} \right)\\
    = \sum_{a, n} \frac{i}{2}\left\{ 
    \bar{\omega}_{\ast a}\left(1-3\eta_a/2\right)\left[\Tau_{mn}^{(a)} X_{1n}^{(a)} \bar{\kappa}_1 - \Eps_{mn}^{(a)}\left(X_{3n}^{(a)} \bar{\kappa}_3 + X_{5n}^{(a)} \bar{\kappa}_5 \right)  \right] \right.\\
    +\bar{\omega}_{\ast a} \eta_a \left[ \Tau_{mn}^{(a)} X_{2n}^{(a)}\bar{\kappa}_1 - \Eps_{mn}^{(a)} \left( X_{4n}^{(a)}\bar{\kappa}_3 + X_{6n}^{(a)}\bar{\kappa}_5 \right) \right]\\
    \left.-\Tau_{mn}^{(a)}X_{1n}^{(a)} \bar{\kappa}_2 + \Eps_{mn}^{(a)}\left( X_{3n}^{(a)} \bar{\kappa}_4 + X_{5n}^{(a)} \bar{\kappa}_6 \right) \right\}.\label{kappa-bar-eigenprob}
\end{multline}
where we define the dimensionless quantities (see also Appendix \ref{X-appx})
\begin{eqnarray}
    \Tau_{mn}^{(a)} = \frac{c_{mn}^{(a)}\wt_a}{n_a T_a},\\%\frac{c_{mn}^{(a)}}{\sqrt{n_a T_a}}\left( \sum_{a^\prime} \frac{n_{a^\prime}}{n_a}\frac{q_{a^\prime}^2}{q_{a}^2}\frac{T_a}{T_{a^\prime}}\right)^{-1/2},\\
    \Eps_{mn}^{(a)} = \frac{c_{mn}^{(a)}\varepsilon_a}{n_a T_a},\\%\frac{c_{mn}^{(a)}}{\sqrt{n_a T_a}}\left( \frac{\beta_a}{b_a}\right)^{1/2},\\
    X^{(a)}_{mn} = \frac{1}{n_{a}}\int d^3v F_{a0} \psi_{ma} \psi_{na}.
\end{eqnarray}
Equation \ref{kappa-bar-eigenprob} is the central result of this paper, a six dimensional algebraic system of equations for unknowns $\bar{\kappa}_i$.  Since the system is homogeneous in these quantities, a non-trivial solution only exists if the determinant (of the matrix of coefficients) vanishes.  This condition determines the eigenvalues $\Lambda$, which, according to Equation \ref{eq:linear-bound}, realize optimal growth of gyrokinetic free energy; see also the discussion in the following section.

Note that a solution of this system, {\em i.e.} $\Lambda$ and $\{\bar{\kappa}_1, \dots, \bar{\kappa}_6\}$, can be substituted into the kinetic expression, Equation \ref{Reduced-kinetic-eigenprob}, to obtain the complete solution for the distribution functions $g_a$.  When the spatial dependence of the solution is taken as $\delta(\ell - \ell_0)$, we obtain a set of orthogonal modes, which can be completed by introducing the null space of \ref{Reduced-kinetic-eigenprob}, namely all distribution functions satisfying $\bar{\kappa}_n = 0$ for all $n$ (including but not limited to those satisfying $\kappa_{n a} = 0$ for all $a$ and $n$).

There are some properties of this system that help make solving it easier.  First, there is the time reversal symmetry of collisionless gyrokinetics.  That is, for any solution $\{\Lambda, \bar{\kappa}_1, \dots, \bar{\kappa}_6\}$, there is another solution $\{-\Lambda, \bar{\kappa}_1^*, \dots, \bar{\kappa}_6^*\}$, as can be seen by taking the complex conjugate of Equation \ref{kappa-bar-eigenprob}, and noting that $\Lambda$ must be real by Hermiticity of \ref{Kinetic-eigenprob}.  Thus there are at most 3 unique non-zero values of $\Lambda^2$ to find.

Additionally, due to the structure of $X_{mn}$ (see Appendix \ref{X-appx}), the linear algebra can be decoupled into a single two-dimensional problem for $\bar{\kappa}_3$ and $\bar{\kappa}_4$, corresponding to perturbations of finite $\delta A_\parallel$, and a four-dimensional problem involving the remaining degrees of freedom, corresponding to mixed perturbations in $\delta \phi$ and $\delta B_\parallel$.  We will denote the positive eigenvalue of the former problem as $\Lambda_3$, and those of the latter problem as $\Lambda_1$ and $\Lambda_2$, reserving $\Lambda_1$ for the electrostatic root (when it is possible to identify one as such).  For symmetry we denote the negative values as $\Lambda_{-n} = -\Lambda_{n}$.

\subsection{Decomposition of energy balance}

Now that we have shown how to calculate the optimal modes, we can show explicitly how they can be used to put the energy balance equation in a pleasing form.  Generalizing the analysis of the schematic system, given by Eqn.\ref{proto-system}, we write the collisionless energy balance equation, \ref{energy-balance}, as

\begin{equation}
    \frac{d}{dt}\sum_{a, \ap} \lang T_a \int d^3 v \frac{g^*_a}{F_{a0}} {\cal H}_{a \ap} g_{\ap} \rang = 2\sum_{a, \ap} \lang T_a \int d^3 v \frac{g^*_a}{F_{a0}} {\cal D}_{a \ap} g_{\ap} \rang
\end{equation}
or, equivalently, using inner-product notation,
\begin{equation}
    \frac{d}{dt}({\bf g}, {\cal H} {\bf g}) = 2({\bf g}, {\cal D} {\bf g}).
\end{equation}
Now using completeness of the eigenmodes we can expand the state as
\begin{equation}
    {\bf g} = \sum_n c_n(\ell) {\bf g}_n,
\end{equation}
noting that there are only 6 eigenmodes of non-zero eigenvalues so the rest of the (infinite) solution space is the null space, {\em i.e.} $\Lambda_n = 0$ for $|n| \geq 4$.  Now taking Eqn.~\ref{Kinetic-eigenprob}, we use orthogonality of eigenmodes of this (generalized) eigenproblem, {\em i.e.} $({\bf g}_m, {\cal H} {\bf g}_n) = 0$  for $m \neq n$ unless $\Lambda_n = \Lambda_m$, and obtain 

\begin{equation}
    \sum_n \frac{d}{dt} \lang |c_n|^2 \rang =  \sum_n 2\Lambda_n \lang|c_n|^2\rang,\label{optimal-energy-balance}  
\end{equation}
where we have normalized the eigenmodes such that $({\bf g}_n, {\cal H} {\bf g}_n) = 1$.  Defining $\Lambda_\mathrm{max} = \max_n \Lambda_n$, we can immediately conclude that

\begin{equation}
    -2 \Lambda_\mathrm{max} \leq \frac{d \ln H}{dt} \leq 2 \Lambda_\mathrm{max}.
\end{equation}
As noted in \citet{helander-plunk-jpp-part-1}, all these equations may be summed over ${\bf k}$ to yield linear and nonlinear bounds on the total free energy of the plasma fluctuations.  In particular, Equation \ref{optimal-energy-balance}, summed over wavenumbers, proves the necessity of instantaneous optimals, solutions with $\Lambda_n > 0$, for the existence of subcritical gyrokinetic turbulence, as noted already by \citet{landreman_plunk_dorland_2015}.

\section{Asymptotic limits}

The general solution of the Equation \ref{kappa-bar-eigenprob}, though easy to obtain numerically, is too lengthy and complicated to gain much insight from when written down in closed form.  We therefore focus on a number of compact asymptotic results for the case of a hydrogen plasma, which also allow comparison with our previously published results \cite{helander-plunk-prl-2021, helander-plunk-jpp-part-1}.

We order all parameters in terms of the small quantity 

\begin{equation}
\epsilon = \sqrt{\frac{b_e}{b_i}} = \sqrt{\frac{m_e T_e}{m_i T_i}} \ll 1.\label{delta-eqn}
\end{equation}
We will take three limits according to the size of the perpendicular wavenumber relative to the Larmor scales, and we will also consider different strengths of the plasma betas, but using the same ordering for the different betas, $\beta_e \sim \beta_i$.  All other dimensionless parameters are assumed order $1$, {\em i.e.} $\eta_i \sim \eta_e \sim \tau \sim 1$, where $\tau = T_i/T_e$ and we assume that $\omega_{*i} \sim \omega_{*e}$ though we need not order these frequencies in terms of $\epsilon$ as this merely translates into an ordering of $\Lambda$, which could be normalized by their amplitude.  In all cases where we numerically evaluate the solutions, we will take $\beta_i = \beta_e = \beta$, $\omega_{*i}=-\omega_{*e} = \omega_{*}$ and $\eta_i = \eta_e = \tau = 1$.

\subsection{Small wavenumber limit: $b_i \sim \epsilon$ and $b_e \sim \epsilon^3$}

In this limit we must be careful to retain first order contributions in the Bessel function expansions because the problem is singular if $b_e = b_i = 0$ is taken identically ($H$ is not positive definite in this case).

For $\Lambda_1$ and $\Lambda_2$, we consider a number of limits on $\beta$.  In the electrostatic limit ($\beta \sim \epsilon^2$) we have

\begin{equation}
    \Lambda _1^2=\frac{\tau  \left(3 (\tau +1) \eta _e^2 \omega _{* e}^2+2 \left(\omega _{* e}-\omega _{* i}\right){}^2+3 \left(\frac{1}{\tau }+1\right) \eta _i^2 \omega _{*
   i}^2\right)}{8 (\tau +1) b_i}.\label{Lambda1-es-limit}
\end{equation}
Note that this result is formally ${\cal O}(\epsilon^{-1})$ and $\Lambda_2 = 0$ at this order.  One can compare this with Eqn.~(6.4) in \citet{helander-plunk-jpp-part-1}, the optimal adiabatic electron result, which has qualitatively different behavior, going to zero with $\eta_i$ and also tending to zero with $k_\alpha$.  This is explained by the fact that the adiabatic electron limit is not obtained as a simple asymptotic limit of the general two-species result, because the ordering and solving of the electron gyrokinetic equation itself is necessary to obtain the adiabatic electron response.  Thus, the two-species result here is not to be taken simply as a more complete result compared with the adiabatic electron result.

\begin{figure}
    \centering
    \includegraphics{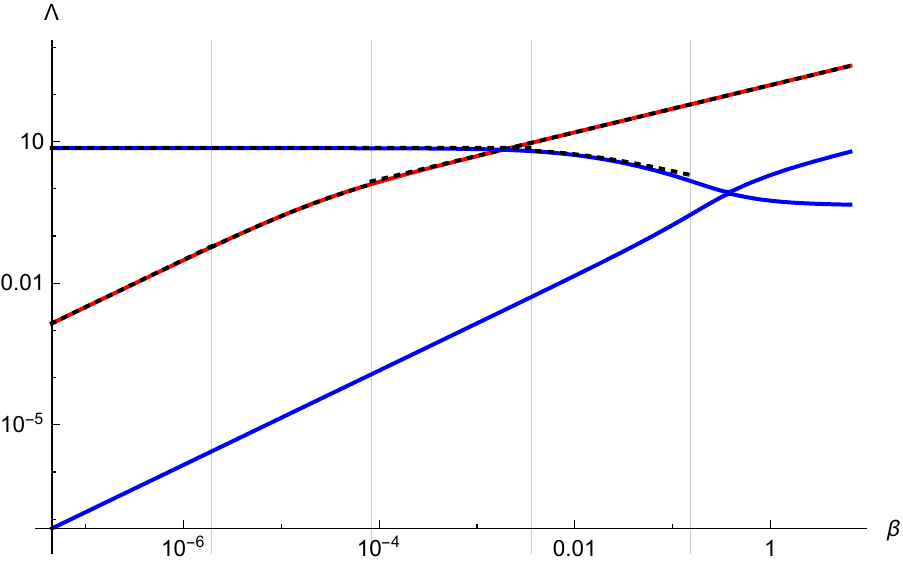}
    \caption{Summary of small wavenumber results.  The numerical results for the mixed modes $\Lambda_1$ and $\Lambda_2$ are plotted in blue, while that for $\Lambda_3$ (the $\delta A_\parallel$ mode) is given in red.  Ranges of $\beta$ (orders of $\epsilon$ ranging from $0$ to $4$), are separated visually by vertical gray lines at intermediate values ($\epsilon^{1/2}$, $\epsilon^{3/2}$, {\em etc}).  The asymptotic results are plotted in dashed-black.  Note that the growth rates are normalized to $|\omega_{i*}|$.}
    \label{fig:small-b}
\end{figure}

Allowing slightly large beta, $\beta \sim \epsilon^1$, the electromagnetic ($\delta B_\parallel$) effects start to mix, making $\Lambda_1$ no longer purely electrostatic:

\begin{equation}
    \Lambda _1^2=\frac{\tau  \left(3 (\tau +1) \eta _e^2 \omega _{* e}^2+2 \left(\omega _{* e}-\omega _{* i}\right){}^2+3 \left(\frac{1}{\tau }+1\right) \eta _i^2 \omega _{*
   i}^2\right)}{4 (\tau +1) \left(2 b_i+\tau  \beta _e+2 \sqrt{\tau  \beta _e \beta _i}+\beta _i\right)}.
\end{equation}
Note that this expression encompasses the $\beta \sim \epsilon^2$ result, Equation \ref{Lambda1-es-limit}.  Note also, that the result can be interpreted as an (initial) finite-$\beta$ stabilization of the electrostatic result; see Figure \ref{fig:small-b}.

For the weakly electromagnetic mode, a single result encompasses all limits of $\beta$ ($\beta_a \sim \epsilon^4$, $\beta_a \sim \epsilon^3$ and $\beta_a \sim \epsilon^2$ and larger):

\begin{equation}
    \Lambda _3^2=\frac{5 \beta _e^2 \eta _e^2 \omega _{* e}^2}{16 b_e \left(2 b_e+\beta _e\right)},\label{Lambda3-small-b}
\end{equation}
where we note that this result also applies to $b_i \sim 1$, as the electron contribution dominates.  The result may be compared with our previous bound, {\em i.e.} the second term in equation (5.6) of \citet{helander-plunk-jpp-part-1}, which can be interpreted as a bound on the electromagnetic contribution to free energy growth.  For general $\eta_e$ one can verify that $\Lambda_3$ of Eqn.~\ref{Lambda3-small-b} is always less than $1/4$ of our previous bound, and goes to zero for $\eta_e = 0$, while our previous bound does not.

A critical value of $\beta$ (of order $\epsilon^2$) can be identified in this (small-$b$) limit, corresponding to the value of $\beta$ above which the root $\Lambda_3$ exceeds the electrostatic root $\Lambda_1$.  An expression for this critical value can be obtained by setting $\Lambda_3 = \Lambda_1$ using Equations \ref{Lambda1-es-limit} and \ref{Lambda3-small-b}, noting that the factor $2 b_e$ can be neglected in the denominator, yielding

\begin{equation}
    \beta_{e,\mathrm{crit}} = \frac{2 \tau  b_e \left(3 (\tau +1) \eta _e^2 \omega _{* e}^2+2 \left(\omega _{* e}-\omega _{* i}\right){}^2+3 \left(\frac{1}{\tau }+1\right) \eta _i^2 \omega _{*
   i}^2\right)}{5 (\tau +1) b_i \eta _e^2 \omega _{* e}^2}.\label{crit-beta-small-b}
\end{equation}
Curiously, we numerically observe that $\Lambda_3$ always seems to be at least as big as the magnetic root $\Lambda_2$ (associated with inclusion of $\delta B_\parallel$), so that the above critical $\beta$ is the one of most relevance for overall stability, and simple asymptotic results like the limits derived here may always be adequate to characterize the actual maximum growth rate of free energy.

To summarize the features of this limit, and confirm the asymptotic results, we numerically solve for roots $\Lambda_1$, $\Lambda_2$ and $\Lambda_3$ and compare them with the values computed from the asymptotic results listed above.  Figure \ref{fig:small-b} shows the result.

\subsection{Intermediate wavenumber limit: $b_i \sim \epsilon^{-1}$, $b_e \sim \epsilon$}

First, we note that for all $\beta$ ($\beta \sim \epsilon^2$, $\beta \sim \epsilon^1$, and $\beta \sim 1$ and larger), the expression given by Equation \ref{Lambda3-small-b} for $\Lambda_3$ still applies in this limit: the ion contribution is still subdominant, and $b_e \ll 1$ also applies here.

For the electrostatic limit ($\beta \sim \epsilon$), we obtain

\begin{multline}
    \Lambda _1^2=\frac{3 \tau ^2 \eta _e^2 \omega _{* e}^2}{8 (\tau +1)}\\ + \frac{\tau  \left(\omega _{* e}^2 \left(6 (\tau +1) \eta _e^2+4\right)+4 \left(\eta _i-2\right) \omega _{* e} \omega _{* i}+\left(5 \eta _i^2-4 \eta
   _i+4\right) \omega _{* i}^2\right)}{16 \sqrt{2 \pi } (\tau +1) \sqrt{b_i}},\label{Lambda1-medium-b-es}
\end{multline}%the expression given 2 equations after Eqn.~(13) of \citet{helander-plunk-prl-2021}, and
which can be compared with our adiabatic-electron electrostatic bound, given by Equation (6.4) of \citet{helander-plunk-jpp-part-1}.  Note that we need to retain the second term, formally smaller than the first by a factor $\epsilon^{1/2}$, for this comparison.  The results can be made comparable by setting $\omega_{*e} = 0$ and additionally taking $\omega_{*i} \rightarrow 0$ while holding $\eta_i \omega_{*i} \sim 1$; see the discussion following Equation \ref{Lambda1-es-limit}.

\begin{figure}
    \centering
    \includegraphics{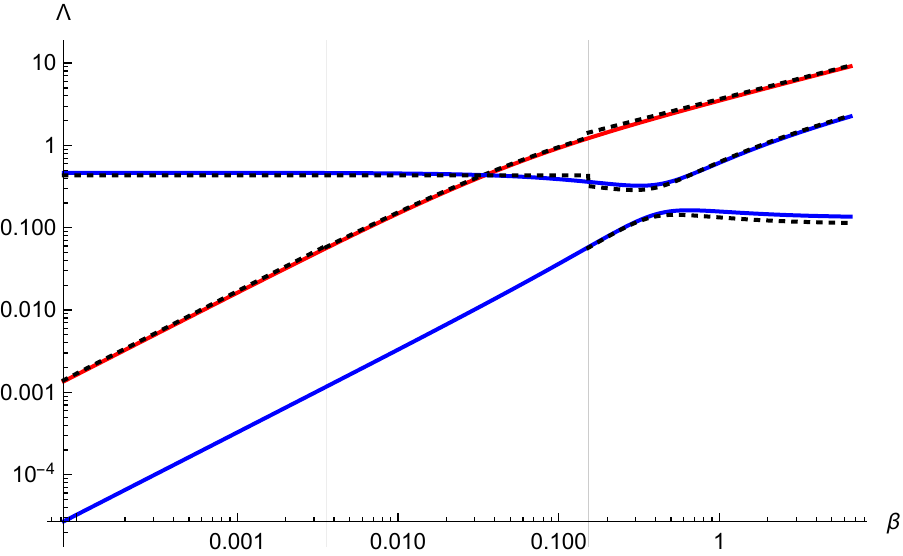}
    \caption{Summary of medium wavenumber results.  The numerical results for the mixed modes $\Lambda_1$ and $\Lambda_2$ are plotted in blue, while that for $\Lambda_3$ (the $\delta A_\parallel$ mode) is given in red.  Ranges of $\beta$ (orders of $\epsilon$ ranging from $0$ to $2$), are separated visually by vertical gray lines at intermediate values ($\epsilon^{1/2}$, $\epsilon^{3/2}$, {\em etc}).  The asymptotic results are plotted in dashed-black.  Note that the growth rates are normalized to $|\omega_{i*}|$.}
    \label{fig:medium-b}
\end{figure}

For $\beta \sim 1$ we obtain (at dominant order) the result
\begin{equation}
    \Lambda _{1,2}^2=\frac{\eta _e^2 \omega _{* e}^2 \left(P\pm \sqrt{P^2 - R}\right)}{16 (\tau +1) \left((\tau +2) \beta _e+2\right)},
\end{equation}
with
\begin{eqnarray}
    P=\left(9 \tau ^2+23 \tau +14\right) \beta _e^2-(9 \tau +10) \tau  \beta _e+6 \tau ^2,\nonumber\\
    R= 18 \tau ^2 (\tau +1) \beta _e^2 \left((\tau +2) \beta _e+2\right).\nonumber
\end{eqnarray}
Note that the dominant (first) term of Equation \ref{Lambda1-medium-b-es} can be recovered as $\beta_e \rightarrow 0$, and that there is also an initial stabilization, as compared with the electrostatic limit associated with finite $\beta_e$, as can be seen in Figure \ref{fig:medium-b}.

We compute the critical $\beta$ as before by equating the asymptotic forms of $\Lambda_1$ and $\Lambda_3$ for the regime where they intersect, $\beta \sim \epsilon^1$ (retaining the first term of Eqn.~\ref{Lambda1-medium-b-es} and the full form of Eqn.~\ref{Lambda3-small-b}):

\begin{equation}
    \beta_{e, \mathrm{crit}} = \frac{\tau b_e \left(\sqrt{9 \tau ^2+60 (\tau +1)}+3 \tau \right) }{5 (\tau +1)},
\end{equation}
where we have selected the positive root to that equation.  Figure \ref{fig:medium-b} summarizes the behavior of the roots across the beta regimes, and confirms the asymptotic results.

\subsection{Large wavenumber limit: $b_e \sim \epsilon^{-1}$, $b_i \sim \epsilon^{-3}$}

For large $b_e$ we obtain finally an asymptotic result for $\Lambda_3$ distinct from the previous.  For $\beta \sim \epsilon^{-1}$ or smaller we obtain,

\begin{equation}
    \Lambda _3^2=\frac{\beta _e^2 \eta _e^2 \omega _{* e}^2}{16 \pi  b_e^3},
\end{equation}
while for $\beta \sim \epsilon^{-2}$ we have

\begin{equation}
    \Lambda _3^2=\frac{\beta _e \eta _e^2 \omega _{* e}^2}{4 \sqrt{2 \pi } b_e^{3/2}}.
\end{equation}

\begin{figure}
    \centering
    \includegraphics{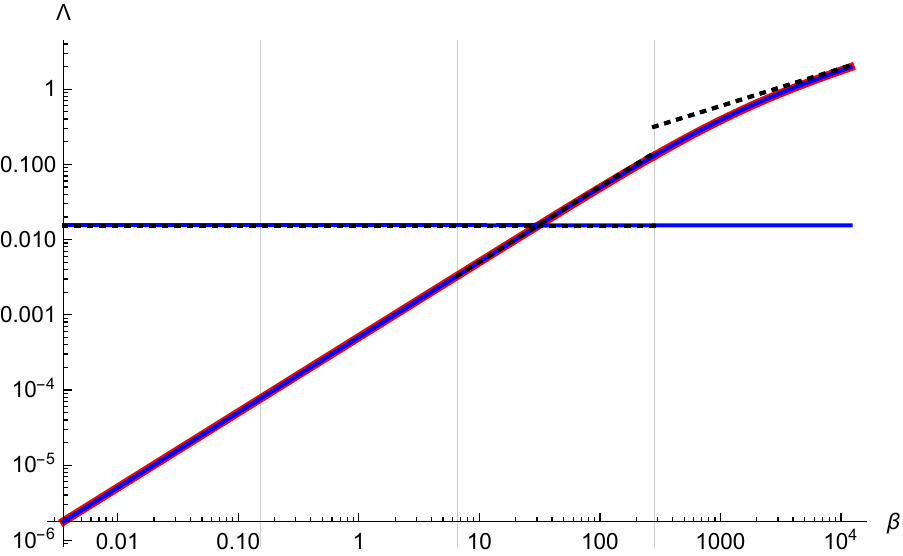}
    \caption{Summary of large wavenumber results.  The numerical results for the mixed modes $\Lambda_1$ and $\Lambda_2$ are plotted in blue, while that for $\Lambda_3$ (the $\delta A_\parallel$ mode) is given in red.  Ranges of $\beta$ (orders of $\epsilon$ ranging from $1$ to $-1$), are separated visually by vertical gray lines at intermediate values ($\epsilon^{1/2}$, $\epsilon^{3/2}$, {\em etc}).  The asymptotic results are plotted in dashed-black.  Note that the growth rates are normalized to $|\omega_{i*}|$.  Note that, curiously, the red curve coincides with one branch of the blue curve, in each range of $\beta$.}
    \label{fig:large-b}
\end{figure}

An electrostatic root is obtained for $\beta \sim 1$, and also survives for $\beta \sim \epsilon^{-1}$ \footnote{As seen in Figure \ref{fig:large-b}, it also seems to be valid for $\beta \sim \epsilon^{-2}$ but we did not prove this.}:

\begin{equation}
    \Lambda _1^2=\frac{\tau ^2 \eta _e^2 \omega _{* e}^2}{8 \pi  (\tau +1)^2 b_e}.
\end{equation}
Note we do not retain higher order contributions for comparison with the adiabatic electron result, which does not discriminate between regimes of $b_e$, since that comparison was already made for the intermediate limit $b_i \sim \epsilon^{-1}$.

For $\beta \sim \epsilon^{-1}$, we find an additional electromagnetic root appears, which, curiously, matches the other electromagnetic root $\Lambda_3$ in this limit

\begin{equation}
    \Lambda _2^2=\frac{\beta _e^2 \eta _e^2 \omega _{* e}^2}{16 \pi  b_e^3}.
\end{equation}

For $\beta \sim \epsilon^{-2}$, the roots continue to match:

\begin{equation}
    \Lambda _2^2=\frac{\beta _e \eta _e^2 \omega _{* e}^2}{4 \sqrt{2 \pi } b_e^{3/2}}.
\end{equation}

The critical value of $\beta$ above which the magnetic roots exceed the electrostatic one is derived as before from the results for $\beta \sim \epsilon^{-1}$:

\begin{equation}
    \beta_{e,\mathrm{crit}} = \frac{\sqrt{2} \tau  b_e}{\tau +1}.
\end{equation}

The results of this limit are summarized in Figure \ref{fig:large-b}.%; note we do not plot the ultra-high $\beta \sim \epsilon^{-2}$ case due to loss of precision in evaluating the Bessel functions for the purely numerical calculation.

\section{Discussion}

Although the essential ideas involved in calculating ``optimal modes'' are simple, the algebra involved is sufficiently complex that a general closed form solution is too lengthy to express in a useful way.  Thus, we have provided a number of compact asymptotic results to demonstrate the essential behavior.

These results compare as expected with the bounds presented in \citet{helander-plunk-prl-2021} and \cite{helander-plunk-jpp-part-1}, {\em i.e.} they are always less than or equal to the previous bounds, and similar qualitatively, though in some special cases much smaller.  The adiabatic electron result of \cite{helander-plunk-jpp-part-1} is a curious case to compare.  One might expect that the two-species electrostatic, long-wavelength limit obtained here might give a similar result, but this is not the case because the adiabatic electron response requires that the electron gyrokinetic equation be ordered and solved at the outset.  Thus the two-species result is not simply a generalization of the one-species adiabatic case.  On the one hand, the two species calculation is able to treat closed-field-line geometries (dipole, Z-pinch), and obtain bounds on the MHD-interchange-like instability there, whose growth tends to a constant as $k_\perp \rightarrow 0$; see for instance \citet{ricci-z-pinch}.  On the other hand, the adiabatic-electron result provides a lower bound, and compares well with expectations of the ITG mode, for which $\gamma \rightarrow 0$ as $k_\perp \rightarrow 0$ \citep{kadomtsev-pogutse, biglari}.  This demonstrates that it is possible to improve, and reduce the complexity of the optimal mode analysis by first applying limits to the fully gyrokinetic system, as should be useful, {\em e.g.}, for the case of trapped electron modes where a bounce-averaged electron response can be used.

We have assumed finite $\delta B_\parallel$ in our analysis, which is chiefly to blame for the added complexity of the algebra, making the results quite general for (local flux-tube) gyrokinetics.  We note that its effect seems mostly subdominant to that of $\delta A_\parallel$ in the sense that at sufficiently low $\beta$ the electrostatic result is dominant, while at large $\beta$, a decoupled mode associated with fluctuations in $\delta A_\parallel$ is dominant, so that the overall maximum growth ($\max_n \Lambda_n$) is never strongly affected by the inclusion of $\delta B_\parallel$.  Thus the weakly electromagnetic result, in which we assume $\delta B_\parallel \approx 0$, may be sufficient to treat many cases of interest, at least if the main goal is establishing overall bounds.

In part I and II in this series, we focus on the Helmholtz free energy, but we note that this is not the only possible choice.  As we have seen, the resulting picture of stability in this case only depends on the strength of certain non-conservative terms in the gyrokinetic equation, and all that survives of the magnetic geometry is contained in the dependence of $k_\perp$ that enters various Bessel functions.  What is lost is the mechanisms of resonance, {\em i.e.} the parallel advection and magnetic drift terms, with the latter being known to explain differences in the stability properties of different kinds of magnetic confinement devices (stellarators, tokamaks, {\em etc.}).  In part III of this series, we will show how the lost effects of magnetic geometry can be recovered by use of a more generalized notion of free energy.

This work has been carried out within the framework of the EUROfusion Consortium, funded by the European Union via the Euratom Research and Training Programme (Grant Agreement No 101052200 — EUROfusion). Views and opinions expressed are however those of the author(s) only and do not necessarily reflect those of the European Union or the European Commission. Neither the European Union nor the European Commission can be held responsible for them.  This work was partly supported by a grant from the Simons Foundation (560651, PH).

\appendix

\section{${\cal H}$ and ${\cal D}$}\label{variation-appx}

Taking Equations \ref{D-eqn} and \ref{H-eqn} and using the field equations, we obtain

\begin{multline}\label{D-eqn-appx}
    D = \sum_{a, b} \frac{i \omega_{\ast a}}{2 n_a n_b} \bigg\langle \int d^3v \int d^3v^\prime g_a^*(v) g_b(v^\prime) \\
    \bigg[\left(1 -\frac{3\eta_a}{2}\right)\left(t_a t_b \psi_{1a}\psi_{1b}^\prime - \varepsilon_a \varepsilon_b \psi_{3a}\psi_{3b}^\prime - \varepsilon_a \varepsilon_b \psi_{5a}\psi_{5b}^\prime \right) \\
    + \eta_a \left( t_a t_b \psi_{2a}\psi_{1b}^\prime - \varepsilon_a \varepsilon_b \psi_{4a} \psi_{3b} - \varepsilon_a \varepsilon_b \psi_{6a} \psi_{5b} \right) \bigg ] \bigg\rangle + \mathrm{c.c.},
\end{multline}
and

\begin{multline}\label{H-eqn-appx}
    H = \sum_{a} \left\langle \int d^3v\frac{T_a |g_a|^2}{F_{a0}} \right\rangle - \sum_{a} \frac{1}{n_a T_a}\left\langle \left| \sum_{b} \frac{1}{n_b}\int d^3v \; t_a t_b \psi_{1b}g_b\right|^2\right\rangle \\
    +\sum_{a,b}\frac{\varepsilon_a\varepsilon_b}{n_a n_b} \left\langle \int d^3v\int d^3v^\prime g_a^*(v) g_b(v^\prime) \left[ \psi_{3a}\psi_{3b}^\prime + \psi_{5a}\psi_{5b}^\prime\right]  \right\rangle + \mathrm{c.c.}
\end{multline}

We define the variation of an arbitrary functional $F$ with respect to the distribution function $g_a$ as

\begin{equation}
    \frac{\delta F}{\delta g_a} = \lim_{\epsilon \rightarrow 0} \frac{d}{d \epsilon} F[g_a + \epsilon h],
\end{equation}
where $h$ is an arbitrary function of phase space variables.  Eqns.~\ref{D-eqn-appx} and \ref{H-eqn-appx} yield

\begin{equation}
    \frac{\delta D}{\delta g_a} = \left\langle \int d^3v \frac{T_a}{F_{a0}} h^* \left\{\sum_{b} {\cal D}_{ab} g_b \right\} \right\rangle + \mathrm{c.c.},
\end{equation}
and

\begin{equation}
    \frac{\delta H}{\delta g_a} = \left\langle \int d^3v \frac{T_a}{F_{a0}} h^* \left\{\sum_{b} {\cal H}_{ab} g_b \right\} \right\rangle + \mathrm{c.c.},
\end{equation}
where the operators ${\cal D}$ and ${\cal H}$ are as defined in Equations \ref{cal-H-eqn} and \ref{cal-D-eqn}.  Note that we have exchanged species labels, and dummy velocity variables of integration to obtain this form.  Because $h$ is arbitrary, these forms substituted into Equation \ref{variational-problem} yield Equation \ref{Kinetic-eigenprob}.
\section{Bessel-type integrals}

To calculate the various integrals involving Bessel functions, we begin with a general form of Weber's integral:

\begin{align}
    {\cal I}_\nu(p,a_1,a_2) &= \int_0^\infty \exp(-p t^2) J_\nu(a_1 t)J_\nu(a_2 t) t dt\nonumber\\
    &= \frac{1}{2p} \exp\left( \frac{-a_1^2-a_2^2}{4p}\right) I_\nu\left(\frac{a_1 a_2}{2p}\right)
\end{align}
where $I_\nu$ is the modified Bessel function of order $\nu$.  The integrals we need to evaluate can be conveniently found in terms of ${\cal I}_\nu$.  We define

\begin{subeqnarray}
    G_{\perp m}(b) &= &2 \int_0^{\infty} x_\perp^{m+1} \exp(-x_\perp^2) J_0^2(\sqrt{2 b} x_\perp^2) dx_\perp,\\
    G_{\perp m}^{(1)}(b) &= &2 \int_0^{\infty} x_\perp^{m+2} \exp(-x_\perp^2) J_0(\sqrt{2 b} x_\perp^2)J_1(\sqrt{2 b} x_\perp^2) dx_\perp,\\
    G_{\perp m}^{(2)}(b) &= &2 \int_0^{\infty} x_\perp^{m+3} \exp(-x_\perp^2) J_1^2(\sqrt{2 b} x_\perp^2) dx_\perp,
\end{subeqnarray}
where $m$ is assumed to be even.  Now we note that these integrals can be evaluated in terms of Weber's integral:

\begin{subeqnarray}
    G_{\perp m}(b) &= &2\left[ \left(-\frac{d}{dp}\right)^{m/2} {\cal I}_0(p,\sqrt{2b},\sqrt{2b})\right]_{p=1},\\
    G_{\perp m}^{(1)}(b) &= &2\left[ \left(-\frac{d}{dp}\right)^{m/2} \left(-\frac{d}{d\lambda}\right) {\cal I}_0(p,\lambda,\sqrt{2b})\right]_{p=1, \lambda = \sqrt{2b}},\\
    G_{\perp m}^{(2)}(b) &= &2\left[ \left(-\frac{d}{dp}\right)^{m/2}\left(-\frac{d}{d\lambda_1}\right)\left(-\frac{d}{d\lambda_2}\right) {\cal I}_0(p,\lambda_1,\lambda_2)\right]_{p=1, \lambda_1 = \lambda_2 = \sqrt{2b}}.
\end{subeqnarray}
The above relations allows us to evaluate the functions

\begin{subeqnarray}
    G_{m,n}(b) &= &G_{\perp m}(b) G_{\parallel n},\\
    G_{m,n}^{(1)}(b) &= &G_{\perp m}^{(1)}(b) G_{\parallel n},\\
    G_{m,n}^{(2)}(b) &= &G_{\perp m}^{(2)}(b) G_{\parallel n}.
\end{subeqnarray}
where 

\begin{equation}
    G_{\parallel n} = \frac{1}{\sqrt{\pi}}\int_{-\infty}^{\infty} \exp{-x_\parallel^2} x_\parallel^n dx_\parallel = \frac{1+(-1)^n}{2\sqrt{\pi}} \Gamma_E\left(\frac{1+n}{2}\right),\\
\end{equation}
and $\Gamma_E$ is the Euler gamma function.

\section{$\mathsfbi{X}(b)$}\label{X-appx}

Bessel functions of various orders enter into the symmetric matrix $\mathsfbi{X}$, written as a function of $b$, depending on the species, whose elements are 

\begin{equation}
    X_{mn}(b_a) = \frac{1}{n_{0a}}\int d^3v F_{a0} \psi_{ma} \psi_{na},
\end{equation}
where $\psi_{m a}$ is the $m$th dimensionless velocity function that appears in the various moments that enter the free energy drive term $D$.  Account for symmetry and oddness over $v_\parallel$ integration, $\mathsfbi{X}$ has the following form:

\begin{equation}
\mathsfbi{X}(b) = \left(
\begin{array}{cccccc}
 X_{11}(b) & X_{12}(b) & 0 & 0 & X_{15}(b) & X_{16}(b) \\
 X_{12}(b) & X_{22}(b) & 0 & 0 & X_{25}(b) & X_{26}(b) \\
 0 & 0 & X_{33}(b) & X_{34}(b) & 0 & 0 \\
 0 & 0 & X_{34}(b) & X_{44}(b) & 0 & 0 \\
 X_{15}(b) & X_{25}(b) & 0 & 0 & X_{55}(b) & X_{56}(b) \\
 X_{16}(b) & X_{26}(b) & 0 & 0 & X_{56}(b) & X_{66}(b) \\
\end{array}
\right)
\end{equation}
The elements $X_{mn}$ are evaluated using the functions $G_{m,n}(b)$, $G_{m,n}^{(2)}(b)$ and $G_{m,n}^{(2)}(b)$, defined in the previous section.  Following convention, we will then evaluate them in terms of the usual gyrokinetic gamma functions

\begin{equation}
    \Gamma_n(b) = \exp(-b)I_n(b)
\end{equation}
Thus, $X_{mn}$ are, for $b$ of arbitrary size, as follows:

\begin{equation}
   X_{11}(b)=G_{0,0}(b)=\Gamma _0(b)
\end{equation}

\begin{equation}
   X_{12}(b)=G_{0,2}(b)+G_{2,0}(b)=\left(\frac{3}{2}-b\right) \Gamma _0(b)+b \Gamma _1(b)
\end{equation}

\begin{equation}
  X_{15}(b)=G^{(1)}_{0,0}(b)=\sqrt{\frac{b}{2}} \left(\Gamma _0(b)-\Gamma \
_1(b)\right)
\end{equation}

\begin{equation}
   X_{16}(b)=G^{(1)}_{0,2}(b)+G^{(1)}_{2,0}(b)=-\frac{\sqrt{b} \left((3 b-5) \Gamma _0(b)+(5-4 b) \Gamma _1(b)+b \Gamma _2(b)\right)}{2 \sqrt{2}}
\end{equation}

\begin{multline}
  X_{22}(b)=G_{0,4}(b)+2 G_{2,2}(b)+G_{4,0}(b)\\
  =\frac{1}{4} \left(\left(6 b^2-20 b+15\right) \Gamma _0(b)+2 b \left((10-4 b) \Gamma _1(b)+b \Gamma _2(b)\right)\right)
\end{multline}

\begin{equation}
  X_{25}(b)=G^{(1)}_{0,2}(b)+G^{(1)}_{2,0}(b) =-\frac{\sqrt{b} \left((3 b-5) \Gamma _0(b)+(5-4 b) \Gamma _1(b)+b \Gamma _2(b)\right)}{2 \sqrt{2}}
\end{equation}

\begin{multline}
    X_{26}(b)=G^{(1)}_{0,4}(b)+2 G^{(1)}_{2,2}(b)+G^{(1)}_{4,0}(b)\\
    =\frac{\sqrt{b}}{4 \sqrt{2}} \left(\left(10 b^2-42 b+35\right) \Gamma _0(b)+\left(-15 b^2+56 b-35\right) \Gamma _1(b)\right)\\
    +\frac{b^{3/2}}{4 \sqrt{2}} \left(2 (3 b-7) \Gamma _2(b)-b \Gamma _3(b)\right)
\end{multline}

\begin{equation}
    X_{33}(b)=G_{0,2}(b)=\frac{\Gamma _0(b)}{2}
\end{equation}

\begin{equation}
    X_{34}(b)=G_{0,4}(b)+G_{2,2}(b)=\frac{1}{4} \left((5-2 b) \Gamma _0(b)+2 b \Gamma _1(b)\right)
\end{equation}

\begin{multline}
    X_{44}(b)=G_{0,6}(b)+2 G_{2,4}(b)+G_{4,2}(b)=\\
    \frac{1}{8} \left(\left(6 b^2-28 b+35\right) \Gamma _0(b)+2 b \left((14-4 b) \Gamma _1(b)+b \Gamma _2(b)\right)\right)
\end{multline}

\begin{equation}
    X_{55}(b)=G^{(2)}_{0,0}(b)=\frac{1}{4} \left(3 b \Gamma _0(b)+(2-4 b) \Gamma _1(b)+b \Gamma _2(b)\right)
\end{equation}

\begin{multline}
    X_{56}(b)=G^{(2)}_{0,2}(b)+G^{(2)}_{2,0}(b)\\
    =\frac{1}{8} \
\left(\left(15 b^2-32 b+10\right) \Gamma _1(b)+(23-10 b) b \Gamma_0(b)+b \left((9-6 b) \Gamma _2(b)+b \Gamma _3(b)\right)\right)
\end{multline}

\begin{multline}
    X_{66}(b)=G^{(2)}_{0,4}(b)+2 G^{(2)}_{2,2}(b)+G^{(2)}_{4,0}(b)\\
    =\frac{1}{16} \left(b \left(35 b^2-188 b+217\right) \Gamma _0(b)+\left(-56 b^3+284 b^2-308 b+70\right) \Gamma _1(b)\right)\\
    +\frac{b}{16} \left(\left(28 b^2-116 b+91\right) \Gamma _2(b)+b \left((20-8 b) \Gamma _3(b)+b \Gamma _4(b)\right)\right)
\end{multline}

\subsection{Asymptotic forms of $X_{mn}$}

For small $b$ we have, to first order, the following forms of $X_{mn}$:

\begin{equation}
\begin{tabular}{l l l l}
 $X_{11}(b)\approx 1-b$    & $X_{12}(b)\approx\frac{3}{2}-\frac{5 b}{2}$ & $X_{15}(b)\approx\frac{\sqrt{b}}{\sqrt{2}}$ & $X_{16}(b)\approx\frac{5 \sqrt{b}}{2 \sqrt{2}}$\\
 $X_{22}(b)\approx\frac{15}{4}-\frac{35 b}{4}$ & $X_{25}(b)\approx\frac{5 \sqrt{b}}{2 \sqrt{2}}$ & $X_{26}(b)\approx\frac{35 \sqrt{b}}{4 \sqrt{2}}$ & $X_{33}(b)\approx\frac{1}{2}-\frac{b}{2}$\\
 $X_{34}(b)\approx\frac{5}{4}-\frac{7 b}{4}$ & $X_{44}(b)\approx\frac{35}{8}-\frac{63 b}{8}$ & $X_{55}(b)\approx b$ & $X_{56}(b)\approx\frac{7 b}{2}$\\
 $X_{66}(b)\approx\frac{63 b}{4}$ & & &
\end{tabular}
\end{equation}

% \begin{equation}
%     X_{11}(b)=1-b
% \end{equation}

% \begin{equation}
%     X_{12}(b)=\frac{3}{2}-\frac{5 b}{2}
% \end{equation}

% \begin{equation}
%     X_{15}(b)=\frac{\sqrt{b}}{\sqrt{2}}
% \end{equation}

% \begin{equation}
%     X_{16}(b)=\frac{5 \sqrt{b}}{2 \sqrt{2}}
% \end{equation}

% \begin{equation}
%     X_{22}(b)=\frac{15}{4}-\frac{35 b}{4}
% \end{equation}

% \begin{equation}
%     X_{25}(b)=\frac{5 \sqrt{b}}{2 \sqrt{2}}
% \end{equation}

% \begin{equation}
%     X_{26}(b)=\frac{35 \sqrt{b}}{4 \sqrt{2}}
% \end{equation}

% \begin{equation}
%     X_{33}(b)=\frac{1}{2}-\frac{b}{2}
% \end{equation}

% \begin{equation}
%     X_{34}(b)=\frac{5}{4}-\frac{7 b}{4}
% \end{equation}

% \begin{equation}
%     X_{44}(b)=\frac{35}{8}-\frac{63 b}{8}
% \end{equation}

% \begin{equation}
%     X_{55}(b)=b,
% \end{equation}

% \begin{equation}
%     X_{56}(b)=\frac{7 b}{2}
% \end{equation}

% \begin{equation}
%     X_{66}(b)=\frac{63 b}{4}
% \end{equation}

For large $b$ we use the following asymptotic forms of $X_{mn}$:

\begin{equation}
\begin{tabular}{l l l l}
 $X_{11}(b)\approx \frac{1}{\sqrt{2 \pi } \sqrt{b}}$    & $X_{12}(b)\approx\frac{1}{\sqrt{2 \pi } \sqrt{b}}$ & $X_{15}(b)\approx\frac{1}{4 \sqrt{\pi } b}$ & $X_{16}(b)\approx\frac{1}{4 \sqrt{\pi } b}$\\
 $X_{22}(b)\approx\sqrt{\frac{2}{\pi b}}$ & $X_{25}(b)\approx\frac{1}{4 \sqrt{\pi } b}$ & $X_{26}(b)\approx\frac{1}{2 \sqrt{\pi } b}$ & $X_{33}(b)\approx\frac{1}{2 \sqrt{2 \pi } \sqrt{b}}$\\
 $X_{34}(b)\approx\frac{1}{\sqrt{2 \pi } \sqrt{b}}$ & $X_{44}(b)\approx\frac{3}{\sqrt{2 \pi } \sqrt{b}}$ & $X_{55}(b)\approx \frac{1}{2 \sqrt{2 \pi } \sqrt{b}}$ & $X_{56}(b)\approx\frac{1}{\sqrt{2 \pi } \sqrt{b}}$\\
 $X_{66}(b)\approx\frac{3}{\sqrt{2 \pi } \sqrt{b}}$ & & &
\end{tabular}
\end{equation}

% \begin{equation}
%     X_{11}(b)=\frac{1}{\sqrt{2 \pi } \sqrt{b}}
% \end{equation}

% \begin{equation}
%     X_{12}(b)=\frac{1}{\sqrt{2 \pi } \sqrt{b}}
% \end{equation}

% \begin{equation}
%     X_{15}(b)=\frac{1}{4 \sqrt{\pi } b}
% \end{equation}

% \begin{equation}
%     X_{16}(b)=\frac{1}{4 \sqrt{\pi } b}
% \end{equation}

% \begin{equation}
%     X_{22}(b)=\frac{\sqrt{\frac{2}{\pi }}}{\sqrt{b}}
% \end{equation}

% \begin{equation}
%     X_{25}(b)=\frac{1}{4 \sqrt{\pi } b}
% \end{equation}

% \begin{equation}
%     X_{26}(b)=\frac{1}{2 \sqrt{\pi } b}
% \end{equation}

% \begin{equation}
%     X_{33}(b)=\frac{1}{2 \sqrt{2 \pi } \sqrt{b}}
% \end{equation}

% \begin{equation}
%     X_{34}(b)=\frac{1}{\sqrt{2 \pi } \sqrt{b}}
% \end{equation}

% \begin{equation}
%     X_{44}(b)=\frac{3}{\sqrt{2 \pi } \sqrt{b}}
% \end{equation}

% \begin{equation}
%     X_{55}(b)=\frac{1}{2 \sqrt{2 \pi } \sqrt{b}}
% \end{equation}

% \begin{equation}
%     X_{56}(b)=\frac{1}{\sqrt{2 \pi } \sqrt{b}}
% \end{equation}

% \begin{equation}
%     X_{66}(b)=\frac{3}{\sqrt{2 \pi } \sqrt{b}}
% \end{equation}

\section{Exact results for the case of a hydrogen plasma}
We give the exact result of optimal growth rates for the electrostatic and weakly electromagnetic cases, defining for compactness the species-dependent elements

\begin{equation}
    X_{mn}^a = X_{mn}(b_a).
\end{equation}

\subsection{Electrostatic limit}

\begin{equation}
\Lambda_1^2 = \frac{C^{(1)}_{ee} \omega _{* e}^2+C^{(1)}_{ee} \omega _{* e} \omega _{* i}+C^{(1)}_{ii} \omega _{* i}^2}{16 (\tau +1) \left(\tau +1-\tau  X_{11}^e-X_{11}^i\right)}
\end{equation}

\begin{multline}
   C^{(1)}_{ii}=\tau  X_{11}^e \left(\left(2-3 \eta _i\right){}^2 X_{11}^i+4 \eta _i \left(\left(2-3 \eta _i\right)
   X_{12}^i+\eta _i X_{22}^i\right)\right)\\
   -4 \eta _i^2 \left((X_{12}^{i})^2-X_{11}^i X_{22}^i\right)
\end{multline}

\begin{multline}
C^{(1)}_{ee}=\tau  X_{11}^e \left(\left(2-3 \eta _e\right){}^2 X_{11}^i+4 \tau  \eta _e^2 X_{22}^e\right)\\
+4 \tau \eta _e \left(X_{11}^i \left(\left(2-3 \eta _e\right)
   X_{12}^e+\eta _e X_{22}^e\right)-\tau  \eta _e (X_{12}^{e})^2\right)
\end{multline}

\begin{equation}
    C^{(1)}_{ie}=-2 \tau  \left(\left(3 \eta _e-2\right) X_{11}^e-2 \eta _e X_{12}^e\right) \left(\left(3 \eta _i-2\right) X_{11}^i-2 \eta _i X_{12}^i\right)
\end{equation}

\subsection{Weakly electromagnetic ``$\delta A_\parallel$ modes''}

\begin{equation}
\Lambda_3^2 = \frac{C^{(3)}_{ee} \omega _{* e}^2+C^{(3)}_{ie} \omega _{* e} \omega _{* i}+C^{(3)}_{ii} \omega _{* i}^2}{16 b_e b_i \left(b_i \beta _e X_{33}^e + b_e \beta _i X_{33}^i + b_e b_i\right)}
\end{equation}

\begin{multline}
   C^{(3)}_{ii}=b_e \beta _i \left(b_i \beta _e X_{33}^e \left(\left(2-3 \eta _i\right){}^2 X_{33}^i+4 \eta _i \left(\left(2-3 \eta _i\right) X_{34}^i+\eta _i
   X_{44}^i\right)\right)\right)\\
   -4 b_e^2 \beta _i^2 \eta _i^2 \left((X_{34}^{i})^2-X_{33}^i X_{44}^i\right)
\end{multline}

\begin{multline}
    C^{(3)}_{ee}=b_i \beta _e \left(X_{33}^e \left(4 b_i \beta _e \eta _e^2 X_{44}^e+b_e \left(2-3 \eta _e\right){}^2 \beta _i X_{33}^i\right)\right)\\
    +4 \eta _e b_i \beta _e\left(\left(b_e \beta _i
   X_{33}^i \left(\left(2-3 \eta _e\right) X_{34}^e+\eta _e X_{44}^e\right)-b_i \beta _e \eta _e (X_{34}^{e})^2\right)\right)
\end{multline}

\begin{equation}
    C^{(3)}_{ie}=-2 b_e b_i \beta _e \beta _i \left(\left(3 \eta _e-2\right) X_{33}^e-2 \eta _e X_{34}^e\right) \left(\left(3 \eta _i-2\right) X_{33}^i-2 \eta _i
   X_{34}^i\right)
\end{equation}

\bibliographystyle{unsrtnat}
\bibliography{energy-bounds-part-2}

\begin{thebibliography}{12}
\providecommand{\natexlab}[1]{#1}
\providecommand{\url}[1]{\texttt{#1}}
\expandafter\ifx\csname urlstyle\endcsname\relax
  \providecommand{\doi}[1]{doi: #1}\else
  \providecommand{\doi}{doi: \begingroup \urlstyle{rm}\Url}\fi

\bibitem[Horton(1999)]{horton-rev}
W.~Horton.
\newblock Drift waves and transport.
\newblock \emph{Rev. Mod. Phys.}, 71:\penalty0 735--778, Apr 1999.
\newblock \doi{10.1103/RevModPhys.71.735}.
\newblock URL \url{https://link.aps.org/doi/10.1103/RevModPhys.71.735}.

\bibitem[Bourdelle et~al.(2015)Bourdelle, Citrin, Baiocchi, Casati, Cottier,
  Garbet, and and]{Bourdelle_2015}
C~Bourdelle, J~Citrin, B~Baiocchi, A~Casati, P~Cottier, X~Garbet, and F~Imbeaux
  and.
\newblock Core turbulent transport in tokamak plasmas: bridging theory and
  experiment with {QuaLiKiz}.
\newblock \emph{Plasma Physics and Controlled Fusion}, 58\penalty0
  (1):\penalty0 014036, dec 2015.
\newblock \doi{10.1088/0741-3335/58/1/014036}.
\newblock URL \url{https://doi.org/10.1088/0741-3335/58/1/014036}.

\bibitem[Hatch et~al.(2016)Hatch, Jenko, Navarro, Bratanov, Terry, and
  Pueschel]{Hatch_2016}
D~R Hatch, F~Jenko, A~Ba{\~{n}}{\'{o}}n Navarro, V~Bratanov, P~W Terry, and M~J
  Pueschel.
\newblock Linear signatures in nonlinear gyrokinetics: {I}nterpreting
  turbulence with pseudospectra.
\newblock \emph{New Journal of Physics}, 18\penalty0 (7):\penalty0 075018, jul
  2016.
\newblock \doi{10.1088/1367-2630/18/7/075018}.
\newblock URL \url{https://doi.org/10.1088/1367-2630/18/7/075018}.

\bibitem[Landreman et~al.(2015)Landreman, Plunk, and
  Dorland]{landreman_plunk_dorland_2015}
Matt Landreman, Gabriel~G. Plunk, and William Dorland.
\newblock Generalized universal instability: transient linear amplification and
  subcritical turbulence.
\newblock \emph{Journal of Plasma Physics}, 81\penalty0 (5):\penalty0
  905810501, 2015.
\newblock \doi{10.1017/S0022377815000495}.

\bibitem[Farrell and Ioannou(1996)]{farrell-generalized-stability-theory}
Brian~F. Farrell and Petros~J. Ioannou.
\newblock Generalized stability theory. {P}art {I}: Autonomous operators.
\newblock \emph{Journal of Atmospheric Sciences}, 53\penalty0 (14):\penalty0
  2025 -- 2040, 1996.
\newblock \doi{10.1175/1520-0469(1996)053<2025:GSTPIA>2.0.CO;2}.

\bibitem[Shankar(1994)]{shankar-book}
Ramamurti Shankar.
\newblock \emph{Principles of quantum mechanics}.
\newblock Plenum Press, New York, 2nd ed edition, 1994.

\bibitem[DelSole(2004)]{del-sole_necessity}
Timothy DelSole.
\newblock The necessity of instantaneous optimals in stationary turbulence.
\newblock \emph{Journal of the Atmospheric Sciences}, 61\penalty0 (9):\penalty0
  1086 -- 1091, 2004.
\newblock \doi{10.1175/1520-0469(2004)061<1086:TNOIOI>2.0.CO;2}.

\bibitem[Helander and Plunk(2022)]{helander-plunk-jpp-part-1}
Per Helander and Gabriel~G. Plunk.
\newblock Modes of optimal growth in gyrokinetics. {P}art {I}. {U}pper bounds.
\newblock \emph{submitted to Journal of Plasma Physics}, 2022.

\bibitem[Helander and Plunk(2021)]{helander-plunk-prl-2021}
P.~Helander and G.~G. Plunk.
\newblock Upper bounds on gyrokinetic instabilities in magnetized plasmas.
\newblock \emph{Phys. Rev. Lett.}, 127:\penalty0 155001, Oct 2021.
\newblock \doi{10.1103/PhysRevLett.127.155001}.
\newblock URL \url{https://link.aps.org/doi/10.1103/PhysRevLett.127.155001}.

\bibitem[Ricci et~al.(2006)Ricci, Rogers, Dorland, and Barnes]{ricci-z-pinch}
Paolo Ricci, B.~N. Rogers, W.~Dorland, and M.~Barnes.
\newblock Gyrokinetic linear theory of the entropy mode in a z pinch.
\newblock \emph{Physics of Plasmas}, 13\penalty0 (6):\penalty0 062102, 2006.
\newblock \doi{10.1063/1.2205830}.
\newblock URL \url{https://doi.org/10.1063/1.2205830}.

\bibitem[Kadomtsev and Pogutse(1970)]{kadomtsev-pogutse}
B.~B. Kadomtsev and O.~P. Pogutse.
\newblock Turbulence in toroidal systems.
\newblock \emph{Rev. Plasmas Phys.}, 5\penalty0 (6):\penalty0 249--400, 1970.

\bibitem[Biglari et~al.(1989)Biglari, Diamond, and Rosenbluth]{biglari}
H.~Biglari, P.~H. Diamond, and M.~N. Rosenbluth.
\newblock Toroidal ion-pressure-gradient-driven drift instabilities and
  transport revisited.
\newblock \emph{Physics of Fluids B: Plasma Physics}, 1\penalty0 (1):\penalty0
  109--118, 1989.
\newblock \doi{10.1063/1.859206}.
\newblock URL \url{http://link.aip.org/link/?PFB/1/109/1}.

\end{thebibliography}

\end{document}